\documentclass[prl,twocolumn]{revtex4-1}

\usepackage{epsfig}
\usepackage{amssymb}
\usepackage{graphicx}
\usepackage{color}
\usepackage{subfigure}

\usepackage{latexsym}

\usepackage{hyperref}

\begin{document}

\newcommand{\nc}{\newcommand}
\newcommand{\rnc}{\renewcommand}



\newcommand{\tcb}{\textcolor{blue}}
\newcommand{\tcr}{\textcolor{red}}
\newcommand{\tcg}{\textcolor{green}}


\def\beq{\begin{equation}}
\def\eeq{\end{equation}}
\def\ba{\begin{array}}
\def\ea{\end{array}}
\def\bea{\begin{eqnarray}}
\def\eea{\end{eqnarray}}
\def\nn{\nonumber}


\def\CMP{Commun. Math. Phys.~}
\def\JHEP{JHEP~}
\def\Pre{Preprint}
\def\PRL{Phys. Rev. Lett.~}
\def\PR {Phys. Rev.~}
\def\CQG {Class. Quant. Grav.~}
\def\PL {Phys. Lett.~}
\def\NP {Nucl. Phys.~}

\def\G{\Gamma}

\def\S{{\bf S}}
\def\C{{\bf C}}
\def\Z{{\bf Z}}
\def\R{{\bf R}}
\def\N{{\bf N}}
\def\M{{\bf M}}
\def\P{{\bf P}}
\def\bm{{\bf m}}
\def\bn{{\bf n}}

\def\CA{{\cal A}}
\def\CB{{\cal B}}
\def\CC{{\cal C}}
\def\CD{{\cal D}}
\def\CE{{\cal E}}
\def\CF{{\cal F}}
\def\CM{{\cal M}}
\def\CG{{\cal G}}
\def\CI{{\cal I}}
\def\CJ{{\cal J}}
\def\CL{{\cal L}}
\def\CK{{\cal K}}
\def\CN{{\cal N}}
\def\CO{{\cal O}}
\def\CP{{\cal P}}
\def\CQ{{\cal Q}}
\def\CR{{\cal R}}
\def\CS{{\cal S}}
\def\CT{{\cal T}}
\def\CV{{\cal V}}
\def\CW{{\cal W}}
\def\CX{{\cal X}}
\def\CY{{\cal Y}}
\def\We{{W_{\mbox{eff}}}}


\newcommand{\p}{\partial}
\newcommand{\bp}{\bar{\partial}}

\newcommand{\half}{\frac{1}{2}}

\newcommand{\bfalpha}{{\mbox{\boldmath $\alpha$}}}
\newcommand{\bfbeta}{{\mbox{\boldmath $\beta$}}}
\newcommand{\bfgamma}{{\mbox{\boldmath $\gamma$}}}
\newcommand{\bfmu}{{\mbox{\boldmath $\mu$}}}
\newcommand{\bfpi}{{\mbox{\boldmath $\pi$}}}
\newcommand{\bfvarpi}{{\mbox{\boldmath $\varpi$}}}
\newcommand{\bftau}{{\mbox{\boldmath $\tau$}}}
\newcommand{\bfeta}{{\mbox{\boldmath $\eta$}}}
\newcommand{\bfxi}{{\mbox{\boldmath $\xi$}}}
\newcommand{\bfkappa}{{\mbox{\boldmath $\kappa$}}}
\newcommand{\bfepsilon}{{\mbox{\boldmath $\epsilon$}}}
\newcommand{\bfTheta}{{\mbox{\boldmath $\Theta$}}}

\newcommand{\bz}{{\bar{z}}}

\newcommand{\dalpha}{\dot{\alpha}}
\newcommand{\dbeta}{\dot{\beta}}
\newcommand{\blambda}{\bar{\lambda}}
\newcommand{\btheta}{{\bar{\theta}}}
\newcommand{\bsigma}{{{\bar{\sigma}}}}
\newcommand{\bepsilon}{{\bar{\epsilon}}}
\newcommand{\bpsi}{{\bar{\psi}}}


\def\ct{\cite}
\def\la{\label}
\def\eq#1{(\ref{#1})}


\def\a{\alpha}
\def\b{\beta}
\def\g{\gamma}
\def\G{\Gamma}
\def\d{\delta}
\def\D{\Delta}
\def\ep{\epsilon}
\def\e{\eta}
\def\ph{\phi}
\def\Ph{\Phi}
\def\ps{\psi}
\def\Ps{\Psi}
\def\k{\kappa}
\def\l{\lambda}
\def\L{\Lambda}
\def\m{\mu}
\def\n{\nu}
\def\th{\theta}
\def\Th{\Theta}
\def\r{\rho}
\def\s{\sigma}
\def\S{\Sigma}
\def\ta{\tau}
\def\o{\omega}
\def\O{\Omega}
\def\pr{\prime}


\def\half{\frac{1}{2}}

\def\goto{\rightarrow}

\def\na{\nabla}
\def\grad{\nabla}
\def\curl{\nabla\times}
\def\div{\nabla\cdot}
\def\pa{\partial}

\def\bra{\left\langle}
\def\ket{\right\rangle}
\def\lb{\left[}
\def\lc{\left\{}
\def\ls{\left(}
\def\lp{\left.}
\def\rp{\right.}
\def\rb{\right]}
\def\rc{\right\}}
\def\rs{\right)}
\def\cl{\mathcal{l}}

\def\vac#1{\mid #1 \rangle}

\def\td#1{\tilde{#1}}
\def\check{ \maltese {\bf Check!}}


\def\Tr{{\rm Tr}\,}
\def\det{{\rm det}\,}


\def\bc#1{\nnindent {\bf $\bullet$ #1} \\ }
\def\ch {$<Check!>$ }
\def\ss {\vspace{1.5cm}}




\preprint{CQUeST--2013-xxxx}


\title{Quasi-Local Conserved Charges in Covariant Theory of Gravity}
\author{Wontae Kim$^{abc}$}\email{wtkim@sogang.ac.kr}
\author{Shailesh Kulkarni$^a$}\email{skulkarnig@gmail.com}
\author{ Sang-Heon Yi$^{a}$}\email{shyi@sogang.ac.kr} 
\affiliation{${}^a$Center for Quantum Spacetime (CQUeST), Sogang University, Seoul 121-742, Korea}
\affiliation{${}^b$Department of Physics,, Sogang University, Seoul 121-742, Korea}
\affiliation{${}^c$ Research Institute for Basic Science,Sogang University, Seoul, 121-742, Republic of Korea}

  







\begin{abstract}
In any generally covariant theory of gravity, we show the relationship between the linearized asymptotically conserved current and its non-linear completion through the identically conserved current. Our formulation for conserved charges is based on the Lagrangian description, and so completely covariant. By using this result, we  give a prescription to define quasi-local conserved charges in any higher derivative gravity.   As applications of our approach, we demonstrate the angular momentum invariance along the radial direction of black holes and reproduce more efficiently the linearized potential on the asymptotic AdS space.



\end{abstract}
\maketitle
\renewcommand{\thefootnote}{\arabic{footnote}}

Identifying the conserved charges  has always remained one of the most important 
and interesting issues in classical general relativity.  There are now various approaches to calculate  conserved quantities, having their own merits and demerits. 
The equivocality among these approaches is the reflection 
of our ineptness to construct the Noether current in an unambiguous way when local gauge symmetries 
are present.  This problem has been tackled, up to certain extent, by focusing on  {\it total}
conserved charges corresponding to Killing vectors at the asymptotic infinity  of the spacetime. 
This is precisely the content of the ADM formalism \cite{Arnowitt:1962hi}. The formalism works well and gives good physical intuition for the asymptotically flat spacetime in Einstein gravity. However, for a generic spacetime, e.g. black holes in AdS spacetime, it is proved to be insufficient. In the another development, initiated by Abbott, Deser and Tekin (ADT)  asymptotic conserved charges are obtained in a covariant manner \cite{Abbott:1981ff, Deser:2002jk}. This approach is especially useful for the asymptotically AdS space-time in Einstein as well as general higher derivative theories of gravity.\\
\indent In contrast,  an appropriate generalization  of asymptotic conserved charges at the quasi-local level is one of unresolved issues in this area. There are several proposals to compute 
quasi-local charges (see~\cite{Szabados:2004vb} for an extensive review and the refs. therein). For instance, the Brown-York method \cite{Brown:1992br} modified by introducing an appropriate counter  term~\cite{Balasubramanian:1999re} has been quite effective.   
However, this construction is not covariant and  it has not been known to be completely consistent with the ADM or ADT formalism~\cite{Abbott:1981ff}. 
Another interesting expression for quasi-local charges is due to  Komar~\cite{Komar:1958wp}. Though this approach is conceptually not so transparent, one may obtain consistent results for total charges in asymptotically flat geometry. However,  the corresponding  expression of mass for AdS black holes gives us a divergent result.  In usual Einstein gravity this difficulty has been cured by subtracting the background contribution. The resultant expression supplemented with the surface term leads to the correct expression for the mass and angular momentum~\cite{Barnich:2003xg}.  This formalism by Barnich {\it et al.} plays a crucial role in determining the central charge in the Kerr/CFT correspondence~\cite{Guica:2008mu}. 
Unfortunately,  there exists no clear-cut procedure for obtaining the
quasi-local expressions when the higher derivative terms are present in the action. This has become one of the major stumbling block in any attempt  to extend the Kerr/CFT duality to generic higher derivative theories of gravity (but, see~\cite{Azeyanagi:2009wf}).  The fact that none of the 
derivations of the quasi-local charges is truly clinching has led to open problems 
leading to alternative approaches with fresh insights.\\
\indent In this letter, we would like to propose a novel way to obtain the quasi-local conserved charges for  black holes in any diffeomorphically invariant theory of gravity.  A new observation is to define the  off-shell Noether potential in terms of the off-shell ({\it i.e.} without using the equations of motion) conserved current. By considering an appropriate variation of the metric,  we 
are able to establish a one-to-one correspondence between the ADT formalism and the linear Noether expressions.  It is worth mentioning  that similar connection has also been  observed  in  Einstein gravity~\cite{Barnich:2001jy, Barnich:2003xg}. However,   the on-shell 
conserved current has been used and thus the whole approach becomes rather complicated.  
On the other hand, in our formalism this complication cease to exist mainly because of the off-shell characteristics  of Noether and ADT potentials.  Once we obtain the linearized Noether 
potential, the quasi-local charges can be easily computed. 

Let us now consider a generally covariant theory of gravity in $D$ spacetime dimensions with an action 
\beq
 I [g_{\mu\nu}] = \frac{1}{16\pi G}\int d^{D}x \sqrt{-g} \, L[g_{\mu\nu}, R, R_{\mu\nu}, \nabla R, \nabla R_{\mu\nu}\cdots]\,.
\eeq
The variation of the above action with respect to  $g^{\mu\nu}$ is given by
\beq 
\delta I =  \frac{1}{16\pi G}\int d^{D}x  \Big[  \sqrt{-g} \CE_{\mu\nu}\delta g^{\mu\nu}+\p_{\mu}\Theta^{\mu}(g;\, \delta g) \Big]\,, \label{genVar}
\eeq
where $\CE^{\mu\nu}=0$ denotes the  equation of motion(EOM) for the metric and $\Theta$ denotes the surface term. 

Under the diffeomorphism $\zeta$, the metric transforms as $\delta g_{\mu\nu} = \nabla_{\mu}\zeta_{\nu} +\nabla_{\nu}\zeta_{\mu}$ while the corresponding change in the  Lagrangian density is  given by $\delta_{\zeta} ( L \sqrt{-g}) =  \p_{\mu} (\sqrt{-g}\, \zeta^{\mu} L)$.
By equating this diffeomorphism transformation with the  generic variation (\ref{genVar}) and exploiting  the Bianchi identity $\nabla_{\mu}\CE^{\mu\nu}=0$, one can obtain the identically conserved (off-shell) current for a generic background metric $g$ as~\cite{Deruelle:2003ps}
\beq \label{OffNoether}
J^{\mu} (g\,; \zeta)= 2\sqrt{-g} \CE^{\mu\nu}(g)\, \zeta_{\nu} + \zeta^{\mu}  \sqrt{-g} L(g)  -  \Theta^{\mu}(g\,; \zeta)\,.  \eeq
Note that this off-shell current becomes the conventional one and leads to  the black hole entropy,  when EOM $\CE=0$ is used~\cite{Wald:1993nt}.
Since $J^{\mu}$ is identically conserved, $\p_{\mu}J^{\mu}=0$,  the anti-symmetric second rank tensor $K^{\mu\nu}$, to which we shall refer as the (off-shell) Noether potential, can be introduced  such that  
$J^{\mu} \equiv \p_{\nu}K^{\mu\nu}$. 
%
%

In order to see the relation of the off-shell current for the diffeomorphism to the linearized conserved current for a Killing vector $\xi$, it is useful to  consider the change in the Noether potential under the variation of the metric $g\rightarrow g_{\mu\nu} + \delta g_{\mu\nu}$ which preserves the Killing vector: $\delta \xi^{\mu} =0$. Then, under such  variation of the metric,  the corresponding change in the off-shell current  can be written  in terms of the Noether potential $K$  as
\beq  \p_{\nu} ( \delta K^{\mu\nu} ) = 2\delta\Big( \sqrt{-g}\, \CE^{\mu\nu} \xi_{\nu} \Big)   + \xi^{\mu}\delta(\sqrt{-g} L) -\delta \Theta^{\mu}(g;\, \xi)\,.    \label{DiffRel}
\eeq

Now, let us go through the construction of conserved charges by Abbott-Deser-Tekin~\cite{Abbott:1981ff,Deser:2002jk}.  For a given solution of EOM with a Killing vector $\xi$, one can introduce a current as $\CJ^{\mu} = \delta \CE^{\mu\nu}\xi_{\nu}$, where $\delta \CE^{\mu\nu}$ denotes the linearization of EOM.  
The current defined in such a way is conserved on-shell.   This
allows us to introduce the so-called on-shell ADT potential, $Q^{\mu\nu}$ as  $\CJ^{\mu}  \equiv \nabla_{\nu} Q^{\mu\nu}$. Note that, unlike the off-shell Noether potential the ADT   potential is defined for the solution of EOM at the linearized level only.  It is precisely this potential that has  been used to compute the ADT charges.

Interestingly, one can also elevate this on-shell current to the off-shell level~\cite{Bouchareb:2007yx} as
\beq  \label{OffADT}
\CJ^{\mu}_{ADT} \equiv  \delta \CE^{\mu\nu}\xi_{\nu} + \CE^{\mu\alpha}h_{\alpha\nu}\xi^{\nu} - \half \xi^{\mu}\CE^{\alpha\beta}h_{\alpha\beta} + \half h \CE^{\mu}_{~ \nu}\xi^{\nu}\,,
\eeq
where  $h$ denotes the variation of the metric, 
$h_{\mu\nu}  \equiv \delta g_{\mu\nu}$,
and indices are raised or lowered by the background metric $g$.  The corresponding off-shell ADT potential $Q^{\mu\nu}_{ADT}$ is given by 
$\CJ^{\mu}_{ADT} = \nabla_{\nu}Q^{\mu\nu}_{ADT}$,
which can be
rewritten in the form of 
\beq 
 \delta (\sqrt{-g}\CE^{\mu\nu}\xi_{\nu} ) - \half \sqrt{-g}\, \xi^{\mu}\CE^{\alpha\beta}h_{\alpha\beta} =  \p_{\nu}(\sqrt{-g}Q^{\mu\nu}_{ADT})\,. \label{offADT}
\eeq
One may note that $\delta \xi^{\mu}=0$ is also assumed in this case, which is the case in the on-shell ADT formalism.

Now, we would like to reveal a relationship between this off-shell ADT potential and the off-shell Noether potential. By combining Eq.~(\ref{offADT}) with  the differential relation~(\ref{DiffRel}) with  $\zeta$ being taken as a Killing vector, and by recalling the generic variation $\delta(\sqrt{-g} L) =  -\sqrt{-g}\CE_{\mu\nu} h^{\mu\nu}+\p_{\mu}\Theta^{\mu}(h)$ from the equation~(\ref{genVar}),
one can see that
%
\beq    \sqrt{-g}\, Q^{\mu\nu}_{ADT} (g\, ; h)=  \frac{1}{2}\,  \delta K^{\mu\nu}(g\,; \xi)     -  \xi^{[\mu} \Theta^{\nu]}(g\, ;  h)\,. \label{MainRes} \eeq
This  is our essential result which shows us explicitly the relationship between the ADT potential and the  Noether potential.  For the final derivation, we have used the covariant form of the $\Theta$ term and the  following configuration space result given  in~\cite{Iyer:1994ys}
\beq  \CL_{\xi}\Theta^{\mu}(g\, ; \delta g)  - \delta \Theta^{\mu}(g\, ;  \xi) =0 \,, \eeq
where $\CL_{\xi}$ denotes the Lie derivative along the Killing vector $\xi$ and $\delta$ does the variation of the background metric $g$. Note that $Q^{\mu\nu}_{ADT}$, itself is conserved  when $g$ and $h$ satisfy the EOM and the linearized EOM, respectively.

We would like to emphasize that  the above derivation is based on the identically conserved or off-shell current in Eq.s~(\ref{OffNoether}) and (\ref{OffADT}).  Consequently,  the background metric does not need to be a solution of  EOM to begin with.  This can be compared  with the configuration space approach taken by Wald {\it et     
al.}~\cite{Wald:1993nt, Iyer:1994ys, Wald:1999wa} wherein the background metric must satisfy the unperturbed Einstein equations. Like in the on-shell case,  the above derivation of the relation between the ADT potential and the Noether potential also suffers from several ambiguities. For instance, since we have used the differentiated relation given in Eq.~(\ref{DiffRel}), the relation can be determined up to an identically vanishing term $\p_{\rho}U^{\mu\nu\rho}$ given by a totally anti-symmetric third rank tensor $U^{\mu\nu\rho}$. However, since the definition of the ADT potential itself contains the same ambiguity or it relies on the cohomology only, one can take the relation between the ADT potential and the variation of the off-shell Noether potential as given in Eq.~(\ref{MainRes}). Furthermore, the surface term $\Theta$ also contains some ambiguities \cite{Wald:1999wa}. Nevertheless, it gives us the correct conserved charges after matching with the linearized charges. This argument can be applied to our approach, too. 
Another point which we would like  to highlight here is that the relation 
(\ref{MainRes}) can be effectively used as a definition for  the ADT potential. 
This way of defining the ADT potential has a distinct advantage over the 
conventional approach when higher derivative terms are present in the theory. 

Usually the linearization  is regarded as meaningful only at the asymptotic infinity to guarantee the validity of the linearization. To overcome this difficulty and to introduce quasi-local conserved charges, it has been proposed to use one parameter path  in the solution space~\cite{Barnich:2003xg, Barnich:2001jy, Wald:1999wa,Barnich:2007bf}.  More explicitly, a path  can be taken as the interpolation through a free parameter $\CQ$ in the solutions of EOM, by the parameter $s$ as  $s\cal{Q}$ $(0 \le s \le 1)$. According to this approach, it is straightforward  to integrate the  linearized potential in Eq.~(\ref{MainRes}) under  the appropriate conditions~\cite{Barnich:2003xg} and to obtain quasi-local conserved charges.  By integrating the one-parameter variable $s$, one can obtain the conserved charge for the on-shell background $g$ as
\beq Q(\xi) = \frac{1}{8\pi G}\int^{1}_{0}ds \int d^{D-2}x_{\mu\nu}\, \sqrt{-g}\, Q^{\mu\nu}_{ADT}(g | s) \,,  \label{Charge}\eeq
where $d^{D-2}x_{\mu\nu}$ denotes the anti-symmetrized integration over coordinates of the co-dimension two subspace.
This expression,  which is the extension of the Einstein gravity case given in~\cite{Barnich:2003xg},   is a meaningful definition for conserved charges  since  $\int ds \sqrt{-g} Q^{\mu\nu}_{ADT}$ is conserved at the non-linear level whenever the integration is well-defined.

The conserved charge given in Eq.~(\ref{Charge}) can be defined in the interior region not just at the asymptotic infinity as in~\cite{Wald:1999wa,Barnich:2007bf}. By using Eq.~(\ref{MainRes}), the quasi-local  conserved charge corresponding to a Killing vector $\xi$ can be eventually written  as
\beq 
 Q(\xi) = \frac{1}{16\pi G}  \int d^{D-2}x_{\mu\nu}
 \Big( \Delta K^{\mu\nu}(\xi)  -   2\xi^{[\mu} \int^{1}_{0}ds~ \Theta^{\nu]}\Big)\,, \label{QuasiCC}
\eeq
where $\Delta K^{\mu\nu}(\xi) \equiv K^{\mu\nu}_{s=1} (\xi)- K^{\mu\nu}_{s=0}(\xi)$ denotes the finite difference between  the two end points of the path   and  $K^{\mu\nu}_{s=0}$ may be taken just as the vacuum solution.
We are going to stress that this can be taken as the proposal for the quasi-local conserved charges in any covariant theory of gravity, which become identical with the ADT charges at the asymptotic infinity.  This  can also be regarded as the generalization of the Einstein gravity case~\cite{Barnich:2007bf} to generally covariant theory of gravity
and as the extension of the result on the asymptotically flat space~\cite{Wald:1999wa} to the arbitrary asymptotic one, especially to the asymptotic AdS one. Note also that our formulation resolves, through the surface term contribution and the matching with the linearized ADT potential,  the normalization issue~\cite{Aros:1999id} of the Noether potential  on the asymptotic AdS space with the background subtraction. 

We now apply our formalism to some specific examples, and  in the process we  obtain some new features.  

\noindent {\it $(i)$ Invariance of angular momentum}:
One of the interesting consequences of our expression given in~(\ref{QuasiCC}) for quasi-local conserved charges is the angular momentum invariance along the radial direction  of black hole solutions. To show this, let us take the rotational Killing vector of axi-symmetric black hole solutions as $\xi_R = \frac{\p}{\p \theta}$ and the integration measure as orthogonal to time and radial directions, $d^{D-2}x_{tr}$. These choices guarantee the absence of the contribution coming from the surface  term in Eq.~(\ref{QuasiCC}). Furthermore, the background value of generalized Noether potential $K^{\mu\nu}_{s=0}$  for the  flat or the AdS space can also be shown to vanish.  As a result, the angular momentum, which  is given by 
\beq 
J_{r} = Q(\xi_R)= \frac{1}{16\pi G}  \int_{\CB_{r}} d^{D-2}x_{\mu\nu}\,  K^{\mu\nu}(\xi_R)\,, \eeq
becomes independent of the radial position, $r$  through the Stokes' theorem. This result explains the validity of the usage of the  Komar's expression for conserved charges in some cases. 
Note that this kind of  invariance is known, in a somewhat different way, in Einstein gravity~\cite{Julia:1998ys}. Moreover, the invariance of the angular momentum has been crucially used  for the matching between the entropies of infra-red and ultra-violate CFT's  
which are dual to the near horizon and asymptotic $AdS_{3}$ spaces of extremal black
  holes~\cite{Kim:2013qra}.

\noindent{\it $(ii)$ Higher curvature case}:
 To obtain the concrete expression for total or quasi-local conserved charges, let us consider a gravity Lagrangian which contains only the invariants of curvature tensors: $L= L[g, R, R^{2},R^{\mu\nu}R_{\mu\nu}, \cdots]$. In this case, it is convenient to treat the curvature tensors as independent variables and  introduce  the covariant current and Noether potential as $J^{\mu} \equiv \sqrt{-g} \CJ^{\mu}$,  $K^{\mu\nu} \equiv \sqrt{-g} \CK^{\mu\nu}$.  Note that $\CE_{\mu\nu}$ can be represented  by
\beq
\CE_{\mu\nu} = P_{(\mu}^{~\alpha\beta\gamma}R_{\nu)\alpha\beta\gamma}   -2\nabla^{\rho}
\nabla^{\sigma}P_{\mu\nu\rho\sigma} -\frac{1}{2}g_{\mu\nu} L\,, \eeq
where $P$-tensor is defined as $P^{\mu\nu\rho\sigma} \equiv  \partial  L / \partial R_{\mu\nu\rho\sigma} $. 
And the surface term $\Theta$ may be taken by
\beq
\Theta^{\mu}(\delta g) = 2\sqrt{-g}[P^{\mu(\alpha\beta)\gamma}\nabla_{\gamma}\delta g_{\alpha\beta} - \delta g_{\alpha\beta}\nabla_{\gamma}P^{\mu(\alpha\beta)\gamma}]\,.
\eeq
Though this surface term has some intrinsic ambiguities, we adopt this form in the following.
Writing the surface term $\Theta$ in term of  
$\delta g_{\mu\nu}$,
one can obtain the above current in the form of
\begin{equation}
\!\CJ^{\mu} = 4\nabla_{\nu}P^{\mu (\rho\sigma) \nu} \nabla_{\rho}\xi_{\sigma}+ 2P^{\mu\nu\rho\sigma}\nabla_{\nu}
\nabla_{\rho}\xi_{\sigma} - 4\xi_{\sigma}\nabla_{\nu}\nabla_{\rho}P^{\mu\nu\rho\sigma} \,.   \end{equation}
Then, the covariant Noether potential
 is given by 
\beq  \CK^{\mu\nu} = 2P^{\mu\nu\rho\sigma}\nabla_{\rho}\xi_{\sigma} - 4\xi_{\sigma}(\nabla_{\rho}P^{\mu\nu\rho\sigma}) \,. \eeq
In Einstein gravity, it turns out that $\CK^{\mu\nu} = 2\nabla^{[\mu}\xi^{\nu]}$. Therefore, the above $\CK^{\mu\nu}$ expression can be regarded as a natural extension from Einstein gravity~\cite{Komar:1958wp,Barnich:2001jy}.  
By using our result and the covariant form of the surface term $\Theta = \sqrt{-g}\bfTheta$, the ADT potential in these theories cab be given by
\beq 2Q^{\mu\nu}_{ADT} = \delta \CK^{\mu\nu} + \frac{1}{2}h\, \CK^{\mu\nu}   -2\xi^{[\mu}\bfTheta^{\nu]}\,. \eeq
Note that  there needs no ad-hoc adjustment  of the famous factor {\it two} in the Komar's expression since it is matched with the twice of the ADT potential  in our formulation. The similar observation was done in the asymptotically flat case in the covariant phase space approach~\cite{Iyer:1994ys}. By using the Noether potential in the higher curvature gravity,   one can obtain  the ADT potential for arbitrary background in a simplified and systematic way. Specifically, one can show that  the ADT potential  for the $R^n$ term is given by
\bea Q^{\mu\nu}_{R^n} &=& nR^{n-1}Q^{\mu\nu}_{R} + n\, \delta R^{n-1} \nabla^{[\mu}\xi^{\nu]}  \nn \\
&&   + 2n\xi^{[\mu}\nabla^{\nu]}\delta R^{n-1}  - n\xi^{[\mu} h^{\nu]}_{~ \alpha}\nabla^{\alpha}R^{n-1}\,,  \nn \eea
where  the ADT potential for the $R$ term, $Q^{\mu\nu}_{R}$,  and $\delta R^{n-1}\equiv  (n-1) R^{n-2}\, \delta R$ term are given respectively by
\bea 
Q^{\mu\nu}_{R} &\equiv & \xi_{\alpha}\nabla^{[\mu}h^{\nu] \alpha} - h^{\alpha [\mu}\nabla_{\alpha} \xi^{\nu]} + \frac{1}{2}h\, \nabla^{[\mu}\xi^{\nu]}  \nn \\
&&   -\xi^{[\mu}\nabla_{\alpha}h^{\nu]\alpha} + \xi^{[\mu}\nabla^{\nu]}h\,, \nn \\
\delta R^{n-1}  &=& (n-1) R^{n-2} ( -R^{\alpha\beta}h_{\alpha\beta} + \nabla^{\alpha}\nabla^{\beta}h_{\alpha\beta} - \nabla^2h )\,. \nn 
\eea
This expression matches completely with the known results for $R$ and $R^2$ terms as special cases. 

\noindent {\it $(iii)$BTZ black holes}:
 In contrast with the angular momentum of AdS black holes, it turns out that, in order to obtain the correct mass of AdS black holes, one needs to subtract the background value of the Noether potential (see some earlier attempt~\cite{Magnon:1985sc}) and should add the surface term contribution.  As an example, let us consider rotating BTZ black hole solutions~\cite{Banados:1992wn} in three-dimensional higher curvature gravity. Since the $P$-tensor is divergence free for the BTZ case, $\nabla_{\mu}P^{\mu\nu\rho\sigma}=0$, 
one can explicitly show that the mass expression  comes entirely from the surface term while  the contribution from $\Delta K$   vanishes.  Furthermore, since the central charge of the boundary dual CFT is proportional to $g_{\mu\nu} (\p L / \p R_{\mu\nu})$,  one can show that   the mass and the angular momentum of BTZ black holes become proportinal to  the  central charge.

In conclusion,  we have established a relation between the ADT potential and the linearized Noether potential  for any higher derivative theory of gravity.  The key ingredient of our derivation was using the off-shell current. This finding has solved the issue raised by Wald {\it et al}. in~\cite{Wald:1999wa}. The equivalence has also enabled us to identify the ADT charges  with the linearized Noether charges, both of which  are defined at the asymptotic 
boundary of the spacetime. By integrating the linearized Noether expression along a one parameter path in the solution space, we have proposed an appropriate definition of quasi-local charges
 which are  matched consistently with the ADT charges. The upshot of our approach is that the Noether potential with a surface term is sufficient
to obtain  quasi-local charges compatible with ADT charges in any covariant 
theory of gravity.   
Since the ADT and Noether  charges are related in one-to-one fashion, we can obtain the ADT potential more easily.  
Our formulation would be very useful in the Kerr/CFT duality.  It would be really interesting to use our result for quasi-local charges  and extend the extremal-Kerr/CFT correspondence to higher derivative theories of 
gravity.\\
\indent{\it Acknowledgements}:
We would like to thank S. Deser and B. Tekin for useful comments on the manuscript. 
This work was supported by the National Research Foundation of Korea(NRF) grant funded by the Korea government(MSIP) through the CQUeST of Sogang University with grant number 2005-0049409. W. Kim was supported  by the Sogang University Research Grant of (2013)201310022. S.-H.Yi was supported by the National Research Foundation of Korea(NRF) grant funded by the Korea government(MOE) (2012R1A1A2004410).


\end{document}